\documentclass[10pt,superscriptaddress,twocolumn,showkeys,aps,prb]{revtex4-1}
\usepackage{graphicx}% Include figure files
\usepackage{dcolumn}% Align table columns on decimal point
\usepackage{bm}% bold math
\usepackage[latin1]{inputenc}
\usepackage{float}
\usepackage{color}
\usepackage{amsmath}
\usepackage{booktabs}

\definecolor{rojo}{rgb}{1,0,0}
\definecolor{verde}{rgb}{0,0.8,0.2}
\definecolor{azul}{rgb}{0,0,1}
\definecolor{rosa}{cmyk}{0,1,0,0}

\usepackage[toc,page]{appendix} 
\usepackage{latexsym}
\usepackage{amsmath}
\usepackage{bm}
\usepackage{amsthm}
\usepackage{bbm}
\usepackage{amsfonts}
\usepackage{amssymb}
\usepackage{epsfig}
\usepackage{hyperref}
\hypersetup{colorlinks=true, citecolor=blue}
\usepackage{soul}

\begin{document}
\title{Ferromagnetic order induced on graphene by Ni/Co proximity effects}
\author{Mayra Peralta}
\affiliation{Centro de F\'isica, Instituto Venezolano de Investigaciones Cient\'ificas, 21827, Caracas, 1020 A, Venezuela.}
\author{Luis Colmenarez}
\affiliation{Departamento de F\'isica, Facultad de Ciencias y Tecnolog\'ia, Universidad de Carabobo, 21827, Caracas, 1020 A, Venezuela.}
\affiliation{Centro de F\'isica, Instituto Venezolano de Investigaciones Cient\'ificas, 21827, Caracas, 1020 A, Venezuela.}
\author{Alejandro L\'opez}
\affiliation{Centro de F\'isica, Instituto Venezolano de Investigaciones Cient\'ificas, 21827, Caracas, 1020 A, Venezuela.}
\author{Bertrand Berche}
\affiliation{Groupe de Physique Statistique, Institut Jean Lamour, Universit\'e de Lorraine, 54506 Vandoeuvre-les-Nancy Cedex, France.}
\affiliation{Centro de F\'isica, Instituto Venezolano de Investigaciones Cient\'ificas, 21827, Caracas, 1020 A, Venezuela.}
\author{Ernesto Medina}
\affiliation{Centro de F\'isica, Instituto Venezolano de Investigaciones Cient\'ificas, 21827, Caracas, 1020 A, Venezuela.}
\affiliation{Groupe de Physique Statistique, Institut Jean Lamour, Universit\'e de Lorraine, 54506 Vandoeuvre-les-Nancy Cedex, France.}
\affiliation{Yachay Tech, School of Physical Sciences \& Nanotechnology, 100119-Urcuqu\'i, Ecuador}

\date{\today}

\begin{abstract}
We build a tight-binding Hamiltonian describing Co/Ni over graphene, contemplating ATOP (a Co/Ni atom on top
of each Carbon atom  of one graphene sublattice) and HCP (one Co/Ni atom per Graphene plaquette) configurations. For the ATOP configuration
the orbitals involved, for the Co/Ni, are the $d_{z^2-r^2}$ which most strongly couples to one graphene sublattice
and the $d_{xz}$, $d_{yz}$ orbitals that couple directly to the second sublattice site. Such configuration is diagonal
in pseudo-spin and spin space, yielding electron doping of the graphene and antiferro-magnetic ordering in the primitive cell in
agreement with DFT calculations. The second, HCP configuration is symmetric in the graphene sublattices 
and only involves coupling to the $d_{xz}$, $d_{yz}$ orbitals. The register of the lattices in this case allows
for a new coupling between nearest neighbour sites, generating non-diagonal terms in the pseudo-spin space and novel spin-kinetic
couplings mimicking a spin-orbit coupling generated by a magnetic coupling. The resulting proximity effect in this case yields ferromagnetic order in the graphene substrate. We derive the band structure in the vicinity of the K points for both configurations, the Bloch wavefunctions and their spin polarization.
\end{abstract}
\maketitle

\section{Introduction}
Graphene-like two dimensional structures have captured the imagination of experimentalists for practical
applications, because of the high hole/electron densities ($10^{13}{\rm cm}^{-2}$ much larger than GaAs electron gas) 
achieved by gating\cite{Novoselov} at a small fraction of the cost and complexity of producing a two dimensional electron gas
with well known semi-conductor technologies. 
One can also build semi-conductors from graphene by breaking the sublattice
A/B symmetry, in systems such as Boron-Nitride\cite{SongBoronNitride}, generating a gap ($\sim 5$ eV) at the K point with a quadratic dispersion.
Because it is undesiderable to introduce substitutional impurities to modify graphene's properties, due to the rapid degradation 
of electron mobilities, one can resort to proximity effects\cite{JapCoGraphene} in order for graphene to inherit 
potentially useful couplings and properties such as a strong Spin-orbit coupling\cite{Marchenko2012}, magnetism\cite{WangProximity} and even chirality\cite{Ostovar}. Deposited transition metals such as Co and 
Ni have matching lattice constants and a few layers easily form on graphene with a large perpendicular magnetic anisotropy\cite{MagneticAnisot}. Such
magnetic layers can be used to introduce new effective couplings between graphene $p_z$ orbitals, thus inducing a strong Rashba type 
coupling, inherited from the Co/Ni overlayer, and also an electron spin polarization with perturbative modifications of the electron mobility.

In this article we consider tight-binding modelling of Co/Ni on planar graphene in two configurations. As found by DFT 
calculations\cite{Macdonald2013}, the lowest energy configuration corresponds to one sub-lattice site of the graphene atop a Co/Ni atom
while the neighbouring sub-lattice atom is in a Co/Ni hcp site. For this configuration, ref.\onlinecite{Macdonald2013} has shown
that the graphene inherits an antiferromagnetic order due mostly to the sublattice asymmetry of the coupling between $p_z$ orbitals
of Carbon and the Co. This type of order lends itself to enhanced
RKKY interactions between Co islands on graphene\cite{VozMediano}, that can be tuned to be either ferro or anti-ferro 
by gate control.

The second configuration we consider corresponds to a global lattice shift
from the previous one where all graphene carbon sites fall at HCP sites of the Co/Ni layer. Referring to the
detailed DFT study of ref.\onlinecite{Macdonald2013} we will first derive the tight-binding model for the first configuration by describing the
orbital overlaps and chemical potentials. We then use such parameterization to estimate the corresponding ones
from the second configuration. While the first configuration yields anti-ferro order on graphene, the second configuration
is symmetric between A/B sublattices and yields ferromagnetic order in the graphene plane.

The summary of this paper is as follows: In section II we describe the two different registers we consider for the 
tight binding model. Focusing first on the ATOP configuration we identify the most salient overlaps involved between the carbon $p_z$ orbitals  and the
$d$ orbitals of Co/Ni (considering only nearest neighbour overlaps), and derive using the band folding scheme,
the effective Hamiltonian for $\pi-$electrons of graphene in the presence of the magnetic overlayer. Next, we shift the lattice register
and consider the HCP configuration, where now the configuration is such that non-diagonal pseudo-spin terms arise,
coupling A and B sublattices. In section III we obtain the Bloch Hamiltonians for both configurations and determine
the band structure and wavefunctions for the bulk samples. The new HCP configuration displays a non-trivial ferromagnetic ordering and a
spin dependent kinetic term proportional to the spin-splitting energy of the Co/Ni covering. Finally, we discuss spin related
properties of the new Hamiltonians for biased and equilibrium current setups.   
\begin{figure}
\includegraphics[scale=0.25]{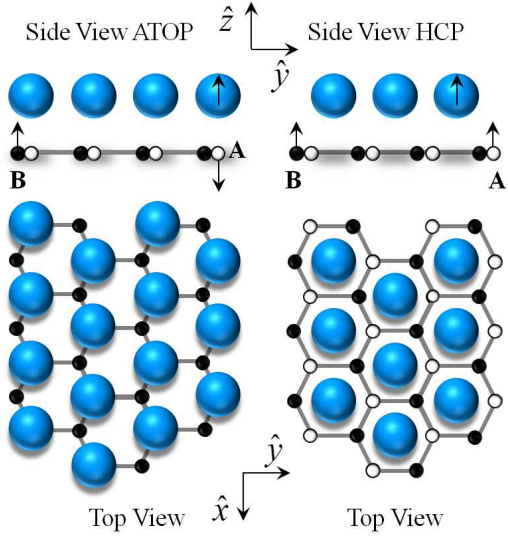}
\caption{Schematic picture of the configurations of a Co monolayer adsorbed on graphene. (Left) ATOP configuration: Co atoms are directly over the atoms of sublattice A and atoms of the sublattice B are in the hcp sites. (Right) HCP configuration: atoms of the sublattices A and B are at hcp sites. In both cases the magnetic order of cobalt, as well as the resulting magnetic order of the sublattices A and B, is indicated.}
\label{Configurations}
\end{figure}
\begin{figure}
\begin{center}
\includegraphics[width=8.5cm]{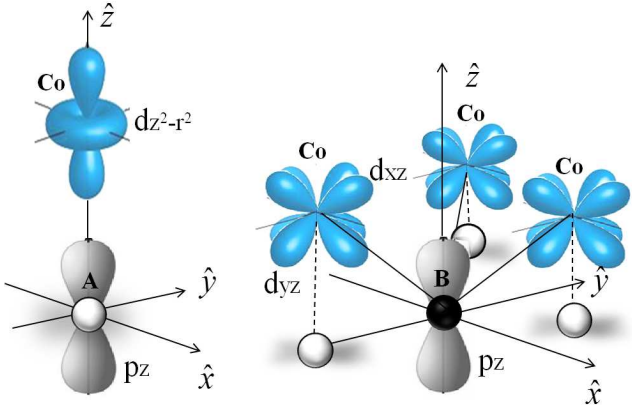}
\end{center}
\caption{Positions of the Co first neighbor around A (left) and B (right), for the ATOP configuration. The orbitals that intervene in the overlaps of Co with A and B are drawn in each case.}
\label{orbitconf1}
\end{figure}

\section{ATOP and HCP configurations and band folding}
Our system consists of a monolayer of Co atoms adsorbed on graphene. A bilayer of Co was shown to be 
stable\cite{Macdonald2013} sustaining a strong anisotropy with magnetization per atom close 
to the bulk values. For the model derived here, we only take into account graphene 
interactions with the first adsorbed layer. We consider two registries for the positions of the Co atoms with 
respect to graphene atoms belonging to the sublattices A and B, which are shown in Fig.~\ref{Configurations}. 
In the configuration of Fig.~\ref{Configurations}(left) (ATOP configuration) the C atoms of the sublattice A are directly under the Co atoms, while atoms of sublattice B are at the hcp sites of the cobalt lattice. In the configuration of Fig.~\ref{Configurations}(right) (HCP configuration), both sublattice atoms are at hcp sites of Co. In the model computations, the first neighbour approximation is used. We will consider that Co is magnetized in the positive 
$\hat{z}$ direction (see Fig.~\ref{Configurations}). {The intrinsic spin-orbit interaction (SOI) and the Rashba coupling 
will not be addressed here since further overlaps will be involved beyond the $d$ orbitals of the Co. Nevertheless, in the absence
of magnetism, the SOI is the only coupling generating spin effects and should be taken into account for interface metals such as Pb\cite{Brey} and
Au\cite{Marchenko2012} to assess e.g. the enhancement of topological properties of graphene. The SOI due to changes in the hybridization of
carbon in graphene as a result of deformations of the surface due to corrugation or hydrogenation\cite{CastroNeto} will be left to future work. }

\subsection{ATOP configuration}

In the ATOP configuration, as shown in the left of Fig.~\ref{orbitconf1}, the orbital $p_{z}$ of a C atom A, $Ap_{z}$, has greatest overlap with the orbital $d_{z^2-r^2}$ (in short $d_{z^2}$), of the Co layer. For the neighbouring B atom (see the right of Fig.~\ref{orbitconf1}), the orbital $p_{z}$, $Bp_{z}$, overlaps preferentially with the orbitals $d_{xz}$ and $d_{yz}$ of the first neighbour Co atoms (there are three Co atoms around each B atom of graphene). The graphene-Co coupled Hamiltonian for the ATOP configuration is the following
\begin{widetext}
\begin{equation}
H_{\rm ATOP}=
\begin{bmatrix}
\begin{matrix}
Ap_{z} & Bp_{z}
\end{matrix} &  \begin{matrix}
                      d_{z^2-r^2} &&&& d_{xz} &&&& d_{yz}
                      \end{matrix}\\
\begin{bmatrix}
    \varepsilon_{p}       & V_{pp\pi}  \\
    V^{*}_{pp\pi}   & \varepsilon_{p}\\
\\
\end{bmatrix}   & \begin{bmatrix}
                        V_{pdz} &&&& 0 &&&& 0 \\
                        0 &&&& \hat{n}_{x} \widetilde{V}_{pd\pi} &&&& \hat{n}_{y} \widetilde{V}_{pd\pi}\\
\\
\end{bmatrix}\\
   \begin{bmatrix}
    -V_{pdz} & 0 \\
    0 &  -\hat{n}_{x} \widetilde{V}_{pd\pi} \\
    0 & -\hat{n}_{y} \widetilde{V}_{pd\pi}
\end{bmatrix} & \begin{bmatrix}
    \varepsilon_{dz^2}+\delta_{1}S_{z} & 0 & 0\\
    0   & \varepsilon_{dxz}+\delta_{2}S_{z}  & 0\\
    0   & 0  & \varepsilon_{dyz}+\delta_{2}S_{z}\\
\end{bmatrix}
\end{bmatrix}.
\label{matrix int1}
\end{equation}
\end{widetext}
where the array consists of four sub-spaces. The upper left sub-space contains the bare $p_z$-orbital site energies of graphene $\varepsilon_{p}$, and the off-diagonal overlaps $V_{pp\pi}$, between A-B sites. The upper right subspace contains the overlaps between the orbitals $Ap_{z}$ and $Bp_{z}$ with the orbitals $d_{z^2}$, $d_{xz}$ and $d_{yz}$ which are $V_{pdz}$, ${\hat{n}_{x}} \widetilde{V}_{pd\pi}$ and ${\hat{n}_{y}} \widetilde{V}_{pd\pi}$ (computed below), respectively, where ${\hat{n}_x,\hat{n}_y}$ are the corresponding direction cosines in a Slater-Koster construction\cite{SlaterKoster,Konschuh}. Note that the
lower left submatrix is the negative of the upper right submatrix since $\langle l'|H|l\rangle=(-1)^{l+l'}\langle l|H|l'\rangle$ (see ref.\onlinecite{Konschuh}) where $l$ is the orbital angular momentum quantum number ($l=1$ for $p$ and $l=2$ for $d$ orbitals).

Finally, the lower right sub-space contains, in the diagonal, the energies of the coupled $d$ orbitals, $\varepsilon_{z^2}$, $\varepsilon_{xz}$ and $\varepsilon_{yz}$.  As we are considering that Co is magnetized in the $\hat{z}$ direction, we add to the $d$ orbital energies the Stoner exchange splittings $\delta_1$ for the orbital $d_{z^2}$ and $\delta_2$ for the orbitals $d_{xz}$ and $d_{yz}$. We assume that $d_{xz}$ and $d_{yz}$ have the same exchange coupling. $S_z$ is the $z$ component Pauli matrix.

The overlaps between orbitals $p_{z}$ and, $d_{xz}$ and $d_{yz}$, are calculated using the relations \cite{SlaterKoster,Konschuh}
\begin{equation}
\label{overlaps1}
\begin{split}
\langle Bp_z|H_{1}|d_{xz}\rangle&=\sqrt{3}n_z^{2}n_xV_{pd\sigma}+n_x(1-2n_z^{2})V_{pd\pi},\\
\langle Bp_z|H_{1}|d_{yz}\rangle&=\sqrt{3}n_z^{2}n_yV_{pd\sigma}+n_y(1-2n_z^{2})V_{pd\pi},
\end{split}
\end{equation}
where $n_{x}$, $n_{y}$ and $n_{z}$ are the direction cosines. Both overlaps have a common factor that only depends on $n_z$, $V_{pd\sigma}$ and $V_{pd\pi}$. In spherical coordinates $n_z=\cos\phi$, where $\phi$ is the polar angle. The first three neighbours, B, of the ATOP site share the same $\phi$ angle. Replacing $n_x=\cos\theta\sin\phi$ and $n_y=\sin\theta\sin\phi$, where $\theta$ is the azimuthal angle in the graphene plane and defining
\begin{equation}
\label{vpz}
\widetilde{V}_{pd\pi}=\sin\phi(\sqrt{3}n_z^{2}V_{pd\sigma}+(1-2n_z^{2})V_{pd\pi}),
\end{equation}
which is a common term for both overlaps, we have
\begin{equation}
\label{overlaps2}
\begin{split}
\langle Bp_z|H_{1}|d_{xz}\rangle&=\hat{n}_x\widetilde{V}_{pd\pi},\\
\langle Bp_z|H_{1}|d_{yz}\rangle&=\hat{n}_y\widetilde{V}_{pd\pi},
\end{split}
\end{equation}
where $\hat{n}_x=\cos\theta$ and $\hat{n}_y=\sin\theta$. 

The eigenvalue equation for Eq.\ref{matrix int1} has the form
\begin{equation}
\left( \begin{array}{cccc}
    H_{\gamma}       & T  \\
    T^{\dagger}   & H_{\chi} \\
\end{array}\right) 
\left (\begin{array}{cc} \gamma \\
\chi
\end{array}\right)= E\left (\begin{array}{cc} \gamma \\
\chi
\end{array}\right),
\end{equation}
with 
\begin{widetext}
\begin{eqnarray}
H_{\gamma}&=&\left(\begin{array}{cccc}
    0       & V_{pp\pi}  \\
    V^{*}_{pp\pi}   & 0\\
\end{array}\right)~;~T=\left (\begin{array}{cccc}
 V_{pdz} & 0 & 0 \\
                        0 & \hat{n}_{x} \widetilde{V}_{pd\pi} & \hat{n}_{y} \widetilde{V}_{pd\pi}\\
\end{array}\right ),\notag \\
H_{\chi}&=&\left(\begin{array}{cccccc}
 (\varepsilon_{dz^2}+\delta_{1}S_{z})-\varepsilon_{p} & 0 & 0\\
    0   &(\varepsilon_{dxz}+\delta_{2}S_{z})-\varepsilon_{p}  & 0\\
    0   & 0  & (\varepsilon_{dyz}+\delta_{2}S_{z})-\varepsilon_{p}\\
\end{array}\right),\notag
\end{eqnarray}
\end{widetext}
where we have taken the energy of the orbital $p_{z}$ of graphene as the reference of zero energy, subtracting $\varepsilon_{p}$ to the diagonals of $H_{\gamma}$ and $H_{\chi}$. The wave function subspaces $\gamma=\left (\psi_{Ap_z},\psi_{Bp_z}\right )$ and $\chi=\left (\psi_{z^2-r^2},\psi_{xz},\psi_{yz}\right)$ are coupled by $T$. 
Eliminating the wavefunction subspace of the Co overlayer ($\chi$) one arrives at
\begin{equation}
\left[ H_{\gamma}+T\left(E-H_{\chi}\right)^{-1}T^{\dagger}\right ]\gamma=E\gamma,
\end{equation} 
where we have ``folded" all the information about the couplings and the Co Hamiltonian into a graphene effective coupling
between A and B sublattices and renormalized the site energies. To linear order in $E$ and lowest order in the coupling $T$, we can expand the inverse
operator so that we obtain
\begin{equation}
\left[ H_{\gamma}-T H^{-1}_{\chi}T^{\dagger}\right ]\gamma\approx E S \gamma,
\end{equation}
where $S=1+T H^{-2}_{\chi}T^{\dagger}$. Now one defines $\Phi=S^{1/2}\gamma$, a function which is normalized $|\Phi|^2\approx \gamma^{\dagger}\gamma+\chi^{\dagger}\chi$
to the same order as the new effective Hamiltonian. The final expression is then
\begin{equation}
S^{-1/2}\left[ H_{\gamma}-T H^{-1}_{\chi}T^{\dagger}\right ]S^{-1/2}\Phi\approx E\Phi.
\end{equation}
The effective Hamiltonian for graphene accounting for its interactions with Co is
\begin{equation}
H_{\rm eff}=S^{-1/2}\left[ H_{\gamma}-T H^{-1}_{\chi}T^{\dagger}\right ]S^{-1/2}.
\label{EffectiveGeneric}
\end{equation}
The inverse of the matrix $H_{\chi}$ is then
\begin{equation}
\label{hchiinv1}
 H_{\chi}^{-1}=\left(
\begin{array}{cccc}
-\frac{(\varepsilon_{p}-\varepsilon_{d})+\delta_{1}S_{z}}{(\varepsilon_{p}-\varepsilon_{d})^{2}-\delta_{1}^{2}} & 0 & 0 \\
0 & -\frac{(\varepsilon_{p}-\varepsilon_{d})+\delta_{2}S_{z}}{(\varepsilon_{p}-\varepsilon_{d})^{2}-\delta_{2}^{2}} & 0 \\
0 & 0 & -\frac{(\varepsilon_{p}-\varepsilon_{d})+\delta_{2}S_{z}}{(\varepsilon_{p}-\varepsilon_{d})^{2}-\delta_{2}^{2}}
\end{array}
\right),
\end{equation}
%\textcolor{magenta}{
%\begin{equation}
%\label{hchiinv1}
% H_{\chi}^{-1}=\frac{1}{\varepsilon_{p}-\varepsilon_{d}}\left(
%\begin{array}{cccc}
% 1+\frac{\delta_{1}S_{z}}{\varepsilon_{p}-\varepsilon_{d}} & 0 & 0 \\
%0 & 1+\frac{\delta_{2}S_{z}}{\varepsilon_{p}-\varepsilon_{d}} & 0 \\
%0 & 0 & 1+\frac{\delta_{2}S_{z}}{\varepsilon_{p}-\varepsilon_{d}}
%\end{array}
%\right).
%\end{equation}

%where we have neglected terms quadratic in the spin-splitting $\delta_{1,2}$ (both negative to favor up spin on Co) in comparison to the difference $(\varepsilon_{p}-\varepsilon_{d})$. 
so the product $TH_{\chi}^{-1}T^{\dag}$ is expanded as
\begin{widetext}
\begin{equation*}
\label{uhchiudag1}
TH_{\chi}^{-1}T^{\dag}= \left(\begin{array}{cccc}
V_{pdz} & 0 & 0 \\
0 & \hat{n}_{mx}\widetilde{V}_{pd\pi} & \hat{n}_{my}\widetilde{V}_{pd\pi}
\end{array}
\right)
\left(
\begin{array}{cccc}
-\frac{(\varepsilon_{p}-\varepsilon_{d})+\delta_{1}S_{z}}{(\varepsilon_{p}-\varepsilon_{d})^{2}-\delta_{1}^{2}} & 0 & 0 \\
0 & -\frac{(\varepsilon_{p}-\varepsilon_{d})+\delta_{2}S_{z}}{(\varepsilon_{p}-\varepsilon_{d})^{2}-\delta_{2}^{2}} & 0 \\
0 & 0 & -\frac{(\varepsilon_{p}-\varepsilon_{d})+\delta_{2}S_{z}}{(\varepsilon_{p}-\varepsilon_{d})^{2}-\delta_{2}^{2}}
\end{array}
\right) 
\left(\begin{array}{cccc}
-V_{pdz} & 0 \\
0 & -\hat{n}_{mx}\widetilde{V}_{pd\pi} \\
0 & -\hat{n}_{my}\widetilde{V}_{pd\pi}
\end{array}
\right),
\end{equation*}
\end{widetext}
where we have included the subindex $m$ to the cosine directors $\hat{n}_{mx,my}$ to indicate that for each of the three Co atoms surrounding a B
type atom on graphene (see Fig.~\ref{orbitconf1}), we have a different overlap.
The product becomes the simple diagonal expression
\begin{eqnarray}
\label{uhchiudag12}
&&T H_{\chi}^{-1}T^{\dag}=\notag \\
&&\left(
\begin{array}{cccc}
\frac{(\varepsilon_{p}-\varepsilon_{d})+\delta_{1}S_{z}}{(\varepsilon_{p}-\varepsilon_{d})^{2}-\delta_{1}^{2}}V_{pdz}^{2} & 0 \\
0 & \frac{(\varepsilon_{p}-\varepsilon_{d})+\delta_{2}S_{z}}{(\varepsilon_{p}-\varepsilon_{d})^{2}-\delta_{2}^{2}}\widetilde{V}_{pd\pi}^{2}[\hat{n}_{mx}^{2} + \hat{n}_{my}^{2}]
\end{array}
\right).\notag \\
\end{eqnarray}

The Hamiltonian for the ATOP configuration is obtained substituting Eq.~(\ref{uhchiudag12}) into Eq.~(\ref{EffectiveGeneric}), approximating $S\sim\mathbbm{1}$ and performing the sum $\sum_{m=1}^{3}[{n}_{mx}^{2} + {n}_{my}^{2}]=3$, which accounts for the contribution to the site energy due to hops of electrons that go from B to Co and return back to B (see ref.\onlinecite{PastawskiMedina}). The effective Hamiltonian is then
\begin{equation}
\label{Heff1}
H_{\rm ATOP}\approx H_{\gamma}-\left(
\begin{array}{cccc}
\frac{(\varepsilon_{p}-\varepsilon_{d})+\delta_{1}S_{z}}{(\varepsilon_{p}-\varepsilon_{d})^{2}-\delta_{1}^{2}}V_{pdz}^{2} & 0 \\
0 & 3\frac{(\varepsilon_{p}-\varepsilon_{d})+\delta_{2}S_{z}}{(\varepsilon_{p}-\varepsilon_{d})^{2}-\delta_{2}^{2}}\widetilde{V}_{pd\pi}^{2}
\end{array}
\right).
\end{equation}
In second quantized form, the Hamiltonian for the full Brillouin zone can be written as
\begin{widetext}
\begin{equation}
\label{hM2}
H_{\rm ATOP}={-\sum_{\langle ij \rangle} \gamma_{0} a_{i}^{\dag}b_{j}} -\frac{(\varepsilon_{p}-\varepsilon_{d})+\delta_{1}S_{z}}{(\varepsilon_{p}-\varepsilon_{d})^{2}-\delta_{1}^{2}}V_{pdz}^{2} \sum_{i}a_{i}^{\dag}a_{i} -3\frac{(\varepsilon_{p}-\varepsilon_{d})+\delta_{2}S_{z}}{(\varepsilon_{p}-\varepsilon_{d})^{2}-\delta_{2}^{2}}\widetilde{V}_{pd\pi}^{2} \sum_{j}b_{j}^{\dag}b_{j},
\end{equation}
\end{widetext}
where {$\gamma_{0}=-V_{pp\pi}$} is the regular off-diagonal kinetic term in graphene and $a_{i}$ and $b_{j}$ are the annihilation operators in the sites A and B graphene sublattices.

%This Hamiltonian can be written in a compact way in terms of the pseudo-spin space Pauli matrix $\sigma_{z}$, the real spin matrix $S_{z}$ and the unitary matrix (either pseudo-spin or spin), as
%\begin{equation}
%\label{Heff12}
%H_{\rm ATOP}=H_{\gamma}+\mu \mathbbm{1}_{\sigma} \mathbbm{1}_{s} - h_{0z} \sigma_{z} \mathbbm{1}_{s} - \frac{h_{z0}}{2}\mathbbm{1}_{\sigma} S_{z} -\frac{h_{zz}}{2} \sigma_{z}S_{z},
%\end{equation}
%where the coefficients $\mu$, $h_{0z}$, $h_{z0}$ and $h_{zz}$ are the same than those obtained in reference \onlinecite{Macdonald2013} by ab initio calculations. Here $\mathbbm{1}_{s}$ and  $\mathbbm{1}_{\sigma}$  are the identity matrices in the spin and pseudo-spin space respectively.

\subsection{HCP configuration}
\begin{figure*}
\begin{center}
\includegraphics[width=11.0cm,height=5.0cm]{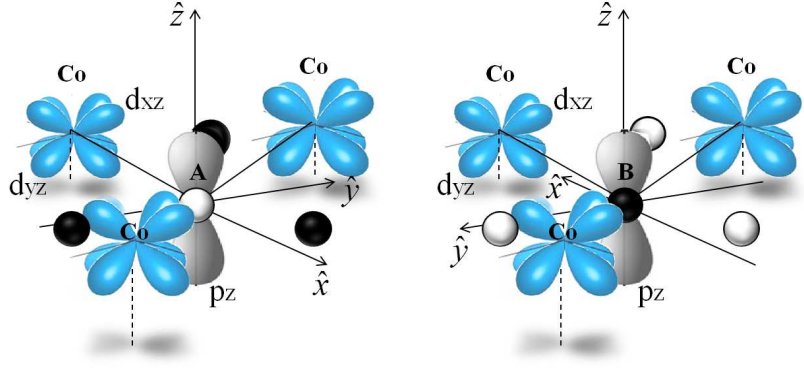}
\end{center}
\caption{ Positions of the Co first neighbor around A (left) and B (right), for the HCP configuration. The orbitals that intervene in the overlaps of Co with A and B are drawn in each case.}
\label{orbitconf2}
\end{figure*}

For the lattice symmetric or HCP configuration, the orbital $d_{z^{2}}$ does not intervene as in the previous case due to the relative positions
of the Co and graphene atoms, as all graphene sites now see the Co/Ni as the B sites in the ATOP configuration. The overlap matrix is given by
\begin{widetext}
\begin{equation}
\begin{bmatrix}
\begin{matrix}
Ap_{z} & Bp_{z}
\end{matrix} &  \begin{matrix}
                      d_{z^2-r^2} &&& d_{xz} &&& d_{yz}
                      \end{matrix}\\
\begin{bmatrix}
    \varepsilon_p       & V_{pp\pi}  \\
    V^{*}_{pp\pi}   & \varepsilon_p \\
\\
\end{bmatrix}   & \begin{bmatrix}
                        0 &&&& \hat{n}_{x} \widetilde{V}_{pd\pi} &&&& \hat{n}_{y} \widetilde{V}_{pd\pi}\\
                        0 &&&& \hat{n}_{x} \widetilde{V}_{pd\pi} &&&& \hat{n}_{y} \widetilde{V}_{pd\pi}\\
\\
\end{bmatrix}\\
   \begin{bmatrix}
    0 & 0 \\
   -\hat{n}_{x} \widetilde{V}_{pd\pi} &  -\hat{n}_{x} \widetilde{V}_{pd\pi} \\
    -\hat{n}_{y} \widetilde{V}_{pd\pi} & -\hat{n}_{y} \widetilde{V}_{pd\pi}
\end{bmatrix} & \begin{bmatrix}
    \varepsilon_{dz^2}+\delta_1S_z   & 0 & 0\\
    0   & \varepsilon_{dxz}+\delta_{2}S_{z}  & 0\\
    0   & 0  & \varepsilon_{dyz}+\delta_{2}S_{z}\\
\end{bmatrix}
\end{bmatrix}.
\label{matrix int2}
\end{equation}
\end{widetext}
In this case A and B see the same environment of three Co atoms at the same distance, as can be seen in Fig.~\ref{orbitconf2}. Both graphene sites interact with them through the orbitals $d_{xz}$ and $d_{yz}$.
\begin{figure}
\begin{center}
\includegraphics[width=5.5cm,height=4.5cm]{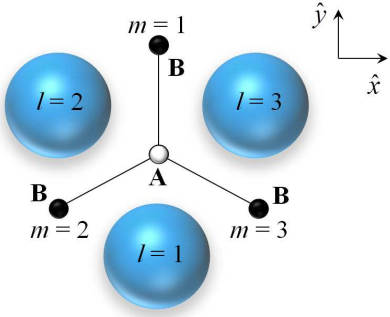}
\end{center}
\caption{ Positions of the Co and B carbon atoms around a carbon atom A, in the HCP configuration of Co/graphene. $l=1,2,3$ represents the Co surrounding A, and $m=1,2,3$ represents the hops from A to B.}
\label{conf2}
\end{figure}

%Inversion of the matrix $H_{\chi}$ gives: 
%\begin{equation}
%\label{hchiinv2}
%H_{\chi}^{-1}=\left(
%\begin{array}{cccc}
%0 & 0 & 0 \\
%0 & \frac{(\varepsilon_{p}-\varepsilon_{d})+\delta_{2}S_{z}}{(\varepsilon_{p}-\varepsilon_{d})^{2}} & 0 \\
%0 & 0 & \frac{(\varepsilon_{p}-\varepsilon_{d})+\delta_{2}S_{z}}{(\varepsilon_{p}-\varepsilon_{d})^{2}}
%\end{array}
%\right).
%\end{equation}
The product $TH_{\chi}^{-1}T^{\dag}$ is now
\begin{widetext}
\begin{equation*}
\label{uhchiudag1}
TH_{\chi}^{-1}T^{\dag}= \left(\begin{array}{cccc}
0 & \hat{n}_{lx}\widetilde{V}_{pd\pi} & \hat{n}_{ly}\widetilde{V}_{pd\pi} \\
0 & \hat{n}_{lmx}\widetilde{V}_{pd\pi} & \hat{n}_{lmy}\widetilde{V}_{pd\pi}
\end{array}
\right)
\left(
\begin{array}{cccc}
-\frac{(\varepsilon_{p}-\varepsilon_{d})+\delta_{1}S_{z}}{(\varepsilon_{p}-\varepsilon_{d})^{2}-\delta_{1}^{2}} & 0 & 0 \\
0 & -\frac{(\varepsilon_{p}-\varepsilon_{d})+\delta_{2}S_{z}}{(\varepsilon_{p}-\varepsilon_{d})^{2}-\delta_{2}^{2}} & 0 \\
0 & 0 & -\frac{(\varepsilon_{p}-\varepsilon_{d})+\delta_{2}S_{z}}{(\varepsilon_{p}-\varepsilon_{d})^{2}-\delta_{2}^{2}}
\end{array}
\right) 
\left(\begin{array}{cccc}
0 & 0 \\
-\hat{n}_{lx}\widetilde{V}_{pd\pi} & -{\hat{n}_{lmx}}\widetilde{V}_{pd\pi} \\
-{\hat{n}_{ly}}\widetilde{V}_{pd\pi} & -\hat{n}_{lmy}\widetilde{V}_{pd\pi}
\end{array}
\right),
\end{equation*}
%we have also neglected terms quadratic in $\delta_{1,2,}$ and
where, as before, $\hat{n}_{l}$ denote direction cosines in the plane that go from the site A to the Co $l=1,2,3$ (A-Co direction) and $\hat{n}_{lm}$ are direction cosines in the plane that go from the site B to the Co $l=1,2,3$ (B-Co direction). Performing the product we have:
\begin{equation}
\label{uhchiudag22}
TH_{\chi}^{-1}T^{\dag}=\left(
\begin{array}{cccc}
\frac{(\varepsilon_{p}-\varepsilon_{d})+\delta_{2}S_{z}}{(\varepsilon_{p}-\varepsilon_{d})^{2}-\delta_{2}^{2}}\widetilde{V}_{pd\pi}^{2}
 [\hat{n}_{lx}^{2} + \hat{n}_{ly}^{2}] & \frac{(\varepsilon_{p}-\varepsilon_{d})+\delta_{2}S_{z}}{(\varepsilon_{p}-\varepsilon_{d})^{2}-\delta_{2}^{2}}\widetilde{V}_{pd\pi}^{2}
[\hat{n}_{lx}\hat{n}_{lmx} + {\hat{n}_{ly}}\hat{n}_{lmy}] \\
\frac{(\varepsilon_{p}-\varepsilon_{d})+\delta_{2}S_{z}}{(\varepsilon_{p}-\varepsilon_{d})^{2}-\delta_{2}^{2}}\widetilde{V}_{pd\pi}^{2}
[\hat{n}_{lx}\hat{n}_{lmx} +{\hat{n}_{ly}}\hat{n}_{lmy}] & \frac{(\varepsilon_{p}-\varepsilon_{d})+\delta_{2}S_{z}}{(\varepsilon_{p}-\varepsilon_{d})^{2}-\delta_{2}^{2}}\widetilde{V}_{pd\pi}^{2}[\hat{n}_{lmx}^{2} + \hat{n}_{lmy}^{2}]
\end{array}
\right).
\end{equation}
\end{widetext}
The Hamiltonian for the symmetric configuration is obtained substituting Eq.~(\ref{uhchiudag22}) into Eq.~(\ref{EffectiveGeneric}), and that $\sum_{l=1}^{3}[{n}_{lx}^{2} + {n}_{ly}^{2}]=3$, which accounts for the contribution to the site energy of hops of electrons that go from A to Co and return back to A. $\sum_{m,l=1}^{3}[{n}_{lmx}^{2} + {n}_{lmy}^{2}]=3$ given that ${n}_{lmx}=\cos\theta'$ and ${n}_{lmy}=\sin\theta'$ where $\theta'$ is the {azimuthal} angle for the overlap between site B and the Co orbitals. The latter sum accounts for the contribution to the site energy of hops of electrons that go from B to Co and return back to B. 

Finally we have the somewhat more complicated summation $\sum_{m,l=1}^{3} [\hat{n}_{lx}\hat{n}_{lmx} + {\hat{n}_{ly}}\hat{n}_{lmy}] $ which is performed in detail in the appendix. The Hamiltonian for this configuration is then
\begin{equation}
\label{Heff2}
H_{\rm HCP}\approx H_{\gamma}-\left(
\begin{array}{cccc}
3\frac{(\varepsilon_{p}-\varepsilon_{d})+\delta_{2}S_{z}}{(\varepsilon_{p}-\varepsilon_{d})^{2}-\delta_{2}^{2}}\widetilde{V}_{pd\pi}^{2} & -\frac{(\varepsilon_{p}-\varepsilon_{d})+\delta_{2}S_{z}}{(\varepsilon_{p}-\varepsilon_{d})^{2}-\delta_{2}^{2}}\widetilde{V}_{pd\pi}^{2}\\
-\frac{(\varepsilon_{p}-\varepsilon_{d})+\delta_{2}S_{z}}{(\varepsilon_{p}-\varepsilon_{d})^{2}-\delta_{2}^{2}}\widetilde{V}_{pd\pi}^{2} & 3\frac{(\varepsilon_{p}-\varepsilon_{d})+\delta_{2}S_{z}}{(\varepsilon_{p}-\varepsilon_{d})^{2}-\delta_{2}^{2}}\widetilde{V}_{pd\pi}^{2}
\end{array}
\right),
\end{equation}
%Using the Pauli, spin, and unitary matrices, we can write the Hamiltonian as:
%\begin{equation}
%\label{Heff22}
%H^{\rm eff}_{\rm Hcp}=H_{\gamma}-\mu' \mathbbm{1}_{\sigma}\mathbbm{1}_{s} - h'_{0x} \sigma_{x}\mathbbm{1}_{s} - h'_{z0}\mathbbm{1}_{\sigma}S_{z} %-h'_{zx} \sigma_{x}S_{z}.
%\end{equation}
%The coefficients $\mu'$, $h'_{0x}$, $h'_{z0}$ and $h'_{zx}$, are the same that the determined by Eq.~(\ref{identconf2}).

%Comparing equations (\ref{hc1}) and (\ref{Heff1}), or (\ref{hM}) and (\ref{Heff12}) for the configuration 1; and (\ref{h2}) and (\ref{Heff2}), or (\ref{h22}) and (\ref{Heff22}) for the configuration 1, we see that we obtained the same Hamiltonians using the two different methods explained in this work. (I need to comment here about the consequences of making the approximation of Eq.~(\ref{effHapr}) over the unitarity of the wave function, but I need to discuss this better).

The complete HCP Hamiltonian, in terms of the creation and annihilation operators in the sites A and B of graphene, $a_{i}$ and $b_{j}$, is given by
\begin{widetext}
\begin{equation}
\label{Heff22}
H_{\rm HCP}=\Big({-\gamma_{0}}+\frac{(\varepsilon_{p}-\varepsilon_{d})+\delta_{2}S_{z}}{(\varepsilon_{p}-\varepsilon_{d})^{2}-\delta_{2}^{2}}\widetilde{V}_{pd\pi}^{2}\Big)\sum_{\langle ij \rangle} a_{i}^{\dag}b_{j} -3\frac{(\varepsilon_{p}-\varepsilon_{d})+\delta_{2}S_{z}}{(\varepsilon_{p}-\varepsilon_{d})^{2}-\delta_{2}^{2}}\widetilde{V}_{pd\pi}^{2} \Big(\sum_{i}a_{i}^{\dag}a_{i} +\sum_{j}b_{j}^{\dag}b_{j}\Big),
\end{equation}
\end{widetext}

%\begin{widetext}
%\begin{equation}
%\label{h2}
 %H_{\rm HCP} = H_{\gamma}-\frac{1}{(\varepsilon_p-\varepsilon_d)}3\widetilde{V}_{pd\pi}^{2} \mathbbm{1}_{\sigma}\mathbbm{1}_{s} -\frac{\delta_2}{(\varepsilon_p-\varepsilon_d)^{2}}3\widetilde{V}_{pd\pi}^{2} \mathbbm{1}_{\sigma}S_{z} +\frac{1}{(\varepsilon_p-\varepsilon_d)}\widetilde{V}_{pd\pi}^{2}  \sigma_{x}\mathbbm{1}_{s}\\  +\frac{\delta_2}{(\varepsilon_p-\varepsilon_d)^{2}}\widetilde{V}_{pd\pi}^{2}\sigma_{x}S_{z},
%\end{equation}
%\end{widetext}
%

%%%%%%%%%%%%%%%%%%%%%%%%%%%%%%%%%%%%%%%
\section{Band structure and magnetic order of the Cobalt/graphene system}
%\begin{figure*}
%\begin{center}
%\includegraphics[width=18.0cm,height=8.3cm]{BandStructures2.eps}
%\end{center}
%\caption{ Band structures in the vicinity of the Dirac point for a monolayer of Co over graphene in: the ATOP configuration (left-down) and the HCP configuration (right-down). The zero is indicated in both plots with a continuous line. In both cases the shift by the chemical potentials $\mu$ and $\mu'$ is observed. The eigenvectors corresponding to each band are indicated for both configurations. $k_{0x}$ is the adimensional wave vector in the $\hat{x}$ direction. In this plot $p_{y}=0$, $p_{x}=\hbar k_{x}-\hbar K_{\xi}$, and  $\xi=+1$. Scheme of the behavior of the total magnetization of graphene $
%\langle S_{z}\rangle_{T}$, given by the contribution of all bands, as a function of the wave vector $k_{x}=k_{0x} (4\pi/3a)$, for: the ATOP configuration (left-up) and the HCP configuration (right-up). The shaded regions at the scheme indicate the zones in which the total magnetization is different from zero. }
%\label{bandstructures}
%\end{figure*}
We now derive the Hamiltonians in reciprocal space in order to determine the band structures of graphene modified by 
adsorbed and polarized Co/Ni in both ATOP and HCP configurations.
The Bloch Hamiltonian is derived by computing the following matrix elements in pseudo-spin space
\begin{eqnarray}
\label{haak}
H_{\rm AA}(\mathbf{k})&=&\frac{1}{N}\sum_{l=1}^{N}\sum_{j=1}^{N} 
e^{i \mathbf{k}\cdot(\mathbf{R}_{Aj}-\mathbf{R}_{Al})} \langle\phi_{Al}| H|\phi_{Aj}\rangle \notag \\
&=&H_{\rm BB}(\mathbf{k})
\end{eqnarray}
where $N$ is the number of unit cells, $\mathbf k$ is the Bloch wavevector and we take $(\mathbf{R}_{Aj}-\mathbf{R}_{Al})=0$ since we do not consider second neighbour interactions (only $j=l$ terms). The off diagonal terms in pseudo-spin space are 
\begin{eqnarray}
\label{habk}
H_{\rm AB}(\mathbf{k})&=&\frac{1}{N}\sum_{l=1}^{N}\sum_{j=1}^{N} e^{i \mathbf{k}\cdot(\mathbf{R}_{Bj}-\mathbf{R}_{Al})} 
\langle\phi_{Al}| H|\phi_{Bj}\rangle\notag \\
&=& H^{\dagger}_{\rm BA}(\mathbf{k}), 
\end{eqnarray}
where $(\mathbf{R}_{Bj}-\mathbf{R}_{Al})=\mathbf{\Delta}_{m}$ is restricted 
to nearest neighbours with $m=1,2,3$ and $\mathbf{\Delta}_1=(0,a/\sqrt{3})$ $\mathbf{\Delta}_2=(a/2,-a/2\sqrt{3})$, $\mathbf{\Delta}_3=(-a/2,-a/2\sqrt{3})$.

For matrix element $H_{\rm AA}$ we consider couplings that connect A to a Co/Ni orbital and then return to same A site (corrections to the site energy see ref.\onlinecite{PastawskiMedina}). In $H_{\rm AB}$ we consider couplings that connect A to Co/Ni sites and then go to one of the three B atoms that are nearest neighbours (corrections to the nearest neighbour matrix elements). In the following we will compute these matrix elements and
derive the resulting band structure in the vicinity of the K points.

\subsection{ATOP Bands}

Using Eqs.~(\ref{haak}) and (\ref{habk}), and evaluating in the vicinity of the K points $K_{\xi}=\xi((4\pi/3a),0)$, the continuum Hamiltonian in reciprocal space for the ATOP configuration can be shown to be
\begin{widetext}
\begin{equation}
\label{Hatopk}
H_{\rm ATOP}(\mathbf{k})= \left(
\begin{array}{cccc}
-\frac{(\varepsilon_{p}-\varepsilon_{d})+\delta_{1}}{(\varepsilon_{p}-\varepsilon_{d})^{2}-\delta_{1}^{2}}V_{pdz}^{2} & v(\xi p_{x}-ip_{y}) & 0 & 0\\
 v(\xi p_{x}+ip_{y}) & -\frac{(\varepsilon_{p}-\varepsilon_{d})-\delta_{1}}{(\varepsilon_{p}-\varepsilon_{d})^{2}-\delta_{1}^{2}}V_{pdz}^{2} & 0 & 0\\
0 & 0 &  -3\frac{(\varepsilon_{p}-\varepsilon_{d})+\delta_{2}}{(\varepsilon_{p}-\varepsilon_{d})^{2}-\delta_{2}^{2}}\widetilde{V}_{pd\pi}^{2} & v(\xi p_{x}-ip_{y}) \\
0 & 0 & v(\xi p_{x}+ip_{y}) & -3\frac{(\varepsilon_{p}-\varepsilon_{d})-\delta_{2}}{(\varepsilon_{p}-\varepsilon_{d})^{2}-\delta_{2}^{2}}\widetilde{V}_{pd\pi}^{2}
\end{array}
\right).
\end{equation}
\end{widetext}
With a more compact parameterization
\begin{widetext}
\begin{equation}
\label{Hatopk2}
H_{\rm ATOP}(\mathbf{k})= \left(
\begin{array}{cccc}
\mu-h_{0z}-\frac{h_{z0}}{2}-\frac{h_{zz}}{2} & v(\xi p_{x}-ip_{y}) & 0 & 0\\
 v(\xi p_{x}+ip_{y}) & \mu-h_{0z}+\frac{h_{z0}}{2}+\frac{h_{zz}}{2} & 0 & 0\\
0 & 0 & \mu+h_{0z}-\frac{h_{z0}}{2}+\frac{h_{zz}}{2} & v(\xi p_{x}-ip_{y}) \\
0 & 0 & v(\xi p_{x}+ip_{y}) & \mu+h_{0z}+\frac{h_{z0}}{2}-\frac{h_{zz}}{2}
\end{array}
\right),
\end{equation}
\end{widetext}
where $\mu=-0.622~$eV, $h_{0z}=0.195~$eV, $h_{z0}=-0.214~$eV and $h_{zz}=-0.766~$eV, are coefficients determined by ab-initio calculations, in the vicinity of the K points, in ref.~\onlinecite{Macdonald2013}. Comparing equations (\ref{Hatopk}) and (\ref{Hatopk2}), we can make the identification
\begin{widetext}
\begin{equation}
\label{identconf1}
  \begin{split}
   -\frac{V_{pdz}^{2}(\varepsilon_p-\varepsilon_d)}{(\varepsilon_p-\varepsilon_d)^{2}-\delta_{1}^{2}} &= \mu-h_{0z},\\
    -3\frac{\widetilde{V}_{pd\pi}^{2}(\varepsilon_p-\varepsilon_d)}{(\varepsilon_p-\varepsilon_d)^{2}-\delta_{2}^{2}} &=\mu+h_{0z},\\
  \end{split}
\quad\quad
  \begin{split}
    \frac{\delta_1V_{pdz}^{2}}{(\varepsilon_p-\varepsilon_d)^{2}-\delta_{1}^{2}} &= \frac{(h_{z0}+h_{zz})}{2},\\
    3\frac{\delta_2\widetilde{V}_{pd\pi}^{2}}{(\varepsilon_p-\varepsilon_d)^{2}-\delta_{2}^{2}} &= \frac{(h_{z0}-h_{zz})}{2}.
   \end{split}
\end{equation}
\end{widetext}
\vspace {2mm}
The identification allows us to determine the coefficients of the Hamiltonian for configurations 
ATOP and later estimate the parameters of model HCP. In the appendix we explicitly write the
coefficients $h_{0z},h_{z0},h_{zz}$ and $\mu$ in terms of the Slater-Koster coefficients. 
%Both $\mu$ and $h_{0z}$ are magnetization independent, while $h_{z0}$ and $h_{zz}$ depend linearly on the magnetization of the Co/Ni overlayer.

\begin{figure*}
\begin{center}
\includegraphics[width=18.0cm,height=8.3cm]{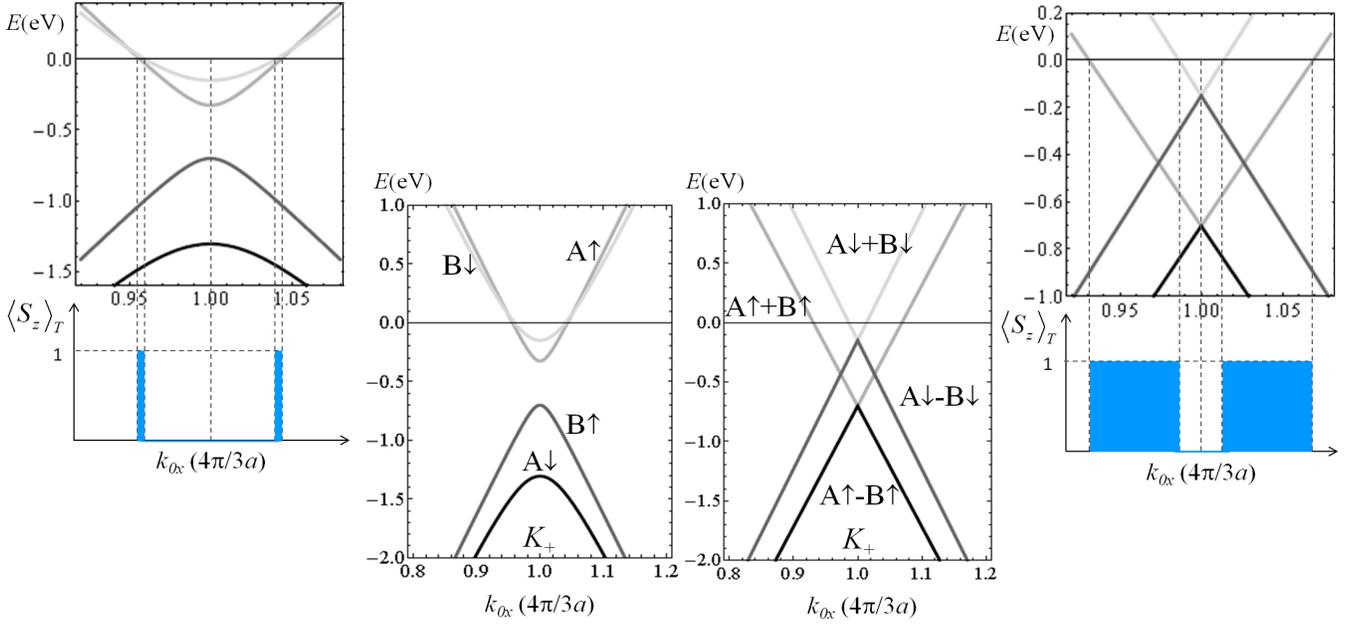}
\end{center}
\caption{ Band structure in the vicinity of the Dirac point for a monolayer of Co over graphene in the ATOP configuration (center-left) and the HCP configuration (center-right). The Fermi level (zero of energy) is indicated in both plots with a continuous line. In both cases the graphene layer is n-doped\cite{Khomyakov}. The eigenvectors corresponding to each band are indicated for both configurations. $k_{0x}$ is the adimensional wave vector in the $\hat{x}$ direction. In this plot $p_{y}=0$, $p_{x}=\hbar k_{x}-\hbar K_{\xi}$, and  $\xi=+1$. The behavior of the total magnetization of graphene 
$\langle S_{z}\rangle_{T}$, given by the contribution of all bands, as a function of the wave vector $k_{x}=k_{0x} (4\pi/3a)$, for the ATOP configuration (upper-left) and the HCP configuration (upper-right). The shaded regions (blue online-only) indicate the zones in which the spin polarization is different from zero.}
\label{bandstructures}
\end{figure*}

In Eq.~(\ref{Hatopk}) {$v=\sqrt{3}a \gamma_0/(2\hbar)$}, with {$-\gamma_0=V_{pp\pi}=-3.033~$}eV. Diagonalization of the Hamiltonian in Eq.~(\ref{Hatopk2}) gives the valence and conduction bands
\begin{eqnarray}
\label{eigenvaluesHatopk}
  \epsilon_v(\mathbf{k})&=&\frac{1}{2}\left (2\mu-s_z h_{z0}-\sqrt{(2h_{0z}+s_z h_{zz})^{2}+4v^{2}\hbar^{2}|\mathbf{k}|^2}\right),\notag \\
  \epsilon_c(\mathbf{k})&=&\frac{1}{2}\left (2\mu-s_z h_{z0}+ \sqrt{(2h_{0z}+s_z h_{zz})^{2}+4v^{2}\hbar^{2}|\mathbf{k}|^2}\right ).\notag \\
\end{eqnarray}
where $s_z=\pm1$ corresponding to the two spin eigenvalues. As a consequence of the A$-$B asymmetry due to the ATOP
geometry, there is a mass term $m_v=(2h_{0z}+s_z h_{zz})/2v$ that will generate spin dependent gaps\cite{AtopGap,Khomyakov} and a quadratic dispersion (see Fig. \ref{bandstructures}). Depending on the material overlaps we can have
a light spin-up holes and heavier spin down holes. On the other hand, for the conduction band it is the up-spin electrons that are lighter in relation to their down spin counterparts.  Something that would be more difficult to assess form DFT studies that is clear from the analytical picture is that an interplay between pseudo-spin and spin active components of the Hamiltonian control the spin dependent effective masses which may have a high contrast making one spin species much more mobile than the other.

Thus the ATOP configuration modifies the linear dispersion in a qualitative way changing the dispersion and mobility of the pristine graphene layer.

For $\mathbf{k}=0$, this Hamiltonian $H_{\rm ATOP}(\mathbf{k})$ is diagonal in the basis
\begin{equation}
\label{baseHeff1}
\left(
\begin{array}{c}
A\uparrow \\
A\downarrow \\
B\uparrow \\
B\downarrow
\end{array}
\right) =\left(
\begin{array}{c}
1 \\
0 \\
0 \\
0
\end{array}
\right), \left(
\begin{array}{c}
0 \\
1 \\
0 \\
0
\end{array}
\right), \left(
\begin{array}{c}
0 \\
0 \\
1 \\
0
\end{array}
\right), \left(
\begin{array}{c}
0 \\
0 \\
0 \\
1
\end{array}
\right).
\end{equation}
Under this condition, we can write:
\begin{eqnarray}
\label{Heff12basis}
H_{\rm ATOP}(\mathbf{k}=0)&=&\mu \mathbbm{1}_{\sigma} \mathbbm{1}_{s} - h_{0z} \sigma_{z} \mathbbm{1}_{s}\notag \\
&-& \frac{h_{z0}}{2}\mathbbm{1}_{\sigma} S_{z} -\frac{h_{zz}}{2} \sigma_{z}S_{z},
\end{eqnarray}
where $\mathbbm{1}_{s}$ and  $\mathbbm{1}_{\sigma}$ are the identity matrices in the spin and pseudo-spin space respectively.

Looking at Eq.~(\ref{Heff12basis}), we can easily recognize the effect of each term on the energy of the system. The first term represents a global energy shift, which is given by the chemical potential $\mu=-0.622~$eV. The negative sign indicates that electrons are transfered from Co to graphene. 
This electron transfer is depicted in Fig.~\ref{bandstructures}(left), where the bare graphene bands are shifted by $\mu$.

The second term is a sub-lattice antisymmetric site energy, $h_{0z}=0.195~$eV. The site energy in this case, decreases in sub-lattice A and increases in B, indicating that the sub-lattice A is more strongly influenced by Co than B (see Figs.~\ref{Configurations} and \ref{orbitconf1}), due to the $p_z-d_{z^2}$ overlap of sublattice A.

The third term is a sub-lattice symmetric spin dependent coupling between Co and graphene. This term gives the spin coupling averaged over sub-lattices A and B. We have $h_{z0}/2=-0.107~$eV, making the states $A\downarrow$ and $B\downarrow$ energetically favorable. {As we have chosen the reference spin magnetization of the Co to be up spin, therefore, the sublattice averaged magnetic order of graphene is antiferromagnetic (AFM) (with respect to Co).}

Finally the fourth term corresponds to a sub-lattice antisymmetric kinetic exchange coupling between Co and graphene spins. We have $h_{zz}/2=-0.383~$eV, which as can be seen in Eq.~(\ref{Heff12basis}), makes the states $A\downarrow$ and $B\uparrow$ energetically favorable, indicating that sublattice A is AFM while sublattice B is FM { with respect to Co spin magnetization.}

Although the previous simple tight binding model seems quite good, there are nevertheless some inconsistencies due to the truncation
of higher order terms involving more complex couplings. We have used a separate spin splitting parameter $\delta_2$ to describe the coupling to 
$d_{xz},d_{yz}$ bands. For there to be a up spin magnetization, the average $\delta$ over all $d$ orbitals of the Co, should be negative. This
is consistent with the top right relation (Eq.\ref{identconf1}) between DFT parameters\cite{Macdonald2013} and tight binding parameters.
Nevertheless, the bottom right equation implies a positive $\delta$ value since $(\varepsilon_p-\varepsilon_d)^{2}>\delta_{2}^{2}$, from
an estimation using Hartree-Fock orbital energies\cite{clementiroetti} and $\delta_2$ from ref.\onlinecite{Macdonald2013}. This cannot be corrected
by including $d_{xy}$ and $d_{x^2-y^2}$ since these contributions add up with the same sign. In the DFT calculation
the ratio between $\delta_1\sim -2.76$eV and $\delta_2\sim -1.083$eV is almost a factor of three but both have consistently a negative sign
i.e. up (majority) spin is lower energy than down (minority) spin.  All the rest of the parameters  of the tight-binding have consistent values
to DFT.

\subsection{HCP bands}
The continuum Hamiltonian in the vicinity of the K points for the HCP configuration is
%\begin{widetext}
%\begin{equation}
%\label{Hhcpk}
%H_{\rm HCP}(\mathbf{k})= \left(
%\begin{array}{cccc}
 %-3\frac{(\varepsilon_{p}-\varepsilon_{d})+\delta_{2}}{(\varepsilon_{p}-\varepsilon_{d})^{2}}\widetilde{V}_{pd\pi}^{2} & 0 & -\frac{v}{\gamma_0} (\gamma_0 + \frac{(\varepsilon_{p}-\varepsilon_{d})+\delta_{2}}{(\varepsilon_{p}-\varepsilon_{d})^{2}}\widetilde{V}_{pd\pi}^{2}) p_{-} & 0\\
 %0 & -3\frac{(\varepsilon_{p}-\varepsilon_{d})-\delta_{2}}{(\varepsilon_{p}-\varepsilon_{d})^{2}}\widetilde{V}_{pd\pi}^{2} & 0 & -\frac{v}{\gamma_0} (\gamma_0 + \frac{(\varepsilon_{p}-\varepsilon_{d})-\delta_{2}}{(\varepsilon_{p}-\varepsilon_{d})^{2}}\widetilde{V}_{pd\pi}^{2})p_{-}\\
 %-\frac{v}{\gamma_0}(\gamma_0 + \frac{(\varepsilon_{p}-\varepsilon_{d})+\delta_{2}}{(\varepsilon_{p}-\varepsilon_{d})^{2}}\widetilde{V}_{pd\pi}^{2}) p_{+} & 0 & -3\frac{(\varepsilon_{p}-\varepsilon_{d})+\delta_{2}}{(\varepsilon_{p}-\varepsilon_{d})^{2}}\widetilde{V}_{pd\pi}^{2} & 0 \\
 %0 & -\frac{v}{\gamma_0}(\gamma_0 + \frac{(\varepsilon_{p}-\varepsilon_{d})-\delta_{2}}{(\varepsilon_{p}-\varepsilon_{d})^{2}}\widetilde{V}_{pd\pi}^{2})p_{+} & 0 & -3\frac{(\varepsilon_{p}-\varepsilon_{d})-\delta_{2}}{(\varepsilon_{p}-\varepsilon_{d})^{2}}\widetilde{V}_{pd\pi}^{2}
%\end{array}
%\right),
%\end{equation}
%\end{widetext}
\begin{widetext}
\begin{equation}
\label{Hhcpk}
H_{\rm HCP}(\mathbf{k})= \left(
\begin{array}{cccc}
-3\frac{\varepsilon+\delta_{2}}{\varepsilon^{2}-\delta_{2}^{2}}\widetilde{V}_{pd\pi}^{2} & 0 & -\frac{v}{\gamma_0} ({-\gamma_0} + \frac{\varepsilon+\delta_{2}}{\varepsilon^{2}-\delta_{2}^{2}}\widetilde{V}_{pd\pi}^{2}) p^{*} & 0\\
0 & -3\frac{\varepsilon-\delta_{2}}{\varepsilon^{2}-\delta_{2}^{2}}\widetilde{V}_{pd\pi}^{2} & 0 & -\frac{v}{\gamma_0} ({-\gamma_0} + \frac{\varepsilon-\delta_{2}}{\varepsilon^{2}-\delta_{2}^{2}}\widetilde{V}_{pd\pi}^{2})p^{*}\\
 -\frac{v}{\gamma_0}({-\gamma_0} + \frac{\varepsilon+\delta_{2}}{\varepsilon^{2}-\delta_{2}^{2}}\widetilde{V}_{pd\pi}^{2}) p & 0 & -3\frac{\varepsilon+\delta_{2}}{\varepsilon^{2}-\delta_{2}^{2}}\widetilde{V}_{pd\pi}^{2} & 0 \\
0 & -\frac{v}{\gamma_0}({-\gamma_0} + \frac{\varepsilon-\delta_{2}}{\varepsilon^{2}-\delta_{2}^{2}} \widetilde{V}_{pd\pi}^{2}) p & 0 & -3\frac{\varepsilon-\delta_{2}}{\varepsilon^{2}-\delta_{2}^{2}}\widetilde{V}_{pd\pi}^{2}
\end{array}
\right),
\end{equation}
\end{widetext}
where $p=\xi p_{x}+ip_{y}$, $p^{*}=\xi p_{x}-ip_{y}$ and $\varepsilon=\varepsilon_{p}-\varepsilon_{d}$. This Hamiltonian can be written as:
%\begin{widetext}
%\begin{equation}
%\label{Hhcpk2}
%H_{\rm HCP}(\mathbf{k})= \left(
%\begin{array}{cccc}
% -\mu'-h'_{z0} & 0 & -\frac{v}{\gamma_0}h_{-}(\xi p_{x}-ip_{y}) & 0\\
% 0 & -\mu'+h'_{z0} & 0 & -\frac{v}{\gamma_0}h_{+}(\xi p_{x}-ip_{y})\\
% -\frac{v}{\gamma_0}h_{-}(\xi p_{x}+ip_{y}) & 0 & -\mu'-h'_{z0} & 0 \\
% 0 & -\frac{v}{\gamma_0}h_{+}(\xi p_{x}+ip_{y}) & 0 & -\mu'+h'_{z0}
%\end{array}
%\right).
%\end{equation}
%\end{widetext}
\begin{widetext}
\begin{equation}
\label{Hhcpk2}
H_{\rm HCP}(\mathbf{k})= \left(
\begin{array}{cccc}
 -\mu'-h'_{z0} & 0 & -\frac{v}{\gamma_0}({-\gamma_{0}}-h'_{0x}-h'_{zx})p^{*} & 0\\
 0 & -\mu'+h'_{z0} & 0 & -\frac{v}{\gamma_0}({-\gamma_{0}}-h'_{0x}+h'_{zx})p^{*}\\
 -\frac{v}{\gamma_0}({-\gamma_{0}}-h'_{0x}-h'_{zx})p & 0 & -\mu'-h'_{z0} & 0 \\
 0 & -\frac{v}{\gamma_0}({-\gamma_{0}}-h'_{0x}+h'_{zx})p & 0 & -\mu'+h'_{z0}
\end{array}
\right).
\end{equation}
\end{widetext}
%where $h_{-}=\gamma_{0}-h'_{0x}-h'_{zx}$ and $h_{+}=\gamma_{0}-h'_{0x}+h'_{zx}$. 
In order to estimate the coefficients $\mu'$, $h_{z0}'$, $h_{0x}'$ and $h_{zx}'$, of this Hamiltonian, we refer to the coupling of the B site in the ATOP configuration which was parameterized by DFT. Comparing equations (\ref{identconf1}), (\ref{Hhcpk}) and (\ref{Hhcpk2}) we arrive at the values
\begin{equation}
\label{identconf2}
  \begin{split}
   \mu'=-(\mu+h_{0z})=0.427~{\rm eV},\\
    h_{z0}'=\frac{h_{z0}-h_{zz}}{2}=0.276~{\rm eV},\\
  \end{split}
\quad
  \begin{split}
   h_{0x}'=-\frac{\mu'}{3}=-0.142~{\rm eV},\\
   h_{zx}'=-\frac{h_{z0}'}{3}=-0.092~{\rm eV}.
  \end{split}
\end{equation}

Diagonalization of the Hamiltonian $H_{\rm HCP}(\mathbf{k})$ gives the eigenvalues
{\begin{equation}
\label{eigenvaluesHhcpk}
  \begin{split}
 \epsilon_v(\mathbf{k})&=-\mu' +s_z h'_{z0}+\frac{v\hbar}{\gamma_0}({-\gamma_{0}}-h'_{0x}+s_z h'_{zx}) |\mathbf{k}|,\\
 \epsilon_c(\mathbf{k})&=-\mu' +s_z h'_{z0}-\frac{v\hbar}{\gamma_0}({-\gamma_{0}}-h'_{0x}+s_z h'_{zx}) |\mathbf{k}|.
\end{split}
\end{equation}}
\vspace{2mm}
where $s_z$ are the eigenvalues of $S_z$ and correspond to the two possible spin orientations. In contrast to the ATOP configuration,
here the dispersion is linear with a modified velocity ${\tilde v}_F={v}({-\gamma_{0}}-h'_{0x}+s_z h'_{zx})/\gamma_0$. Of course, corrections to velocities are one order of magnitude smaller that the pristine graphene values (see Eqs.~\ref{identconf2}).

One can diagonalize the Hamiltonian $H_{\rm HCP}(\mathbf{k})$ in the basis
\begin{widetext}
\begin{eqnarray}
\label{baseHeff2}
\left(
\begin{array}{c}
A\uparrow \\
A\downarrow \\
B\uparrow \\
B\downarrow
\end{array}
\right) &=&\frac{1}{\sqrt{2}}\left(
\begin{array}{c}
0 \\
1 \\
0 \\
\xi e^{i\xi \phi_{k}}
\end{array}
\right), \frac{1}{\sqrt{2}} \left(
\begin{array}{c}
1 \\
0 \\
\xi e^{i\xi \phi_{k}} \\
0
\end{array}
\right), %\notag \\
%&&
\frac{1}{\sqrt{2}} \left(
\begin{array}{c}
0 \\
1 \\
0 \\
-\xi e^{i\xi \phi_{k}}
\end{array}
\right), \frac{1}{\sqrt{2}} \left(
\begin{array}{c}
1 \\
0 \\
-\xi e^{i\xi \phi_{k}} \\
0
\end{array}
\right),
\end{eqnarray}
\end{widetext}
where $\phi_{k}=\arctan(p_{y}/p_{x})$. Given that in Eq.~(\ref{Hhcpk}) we have other non diagonal terms besides the bare graphene terms, and therefore we have other $\mathbf{k}$ dependent terms, we cannot make a useful analysis at $\mathbf{k}=0$ as in the ATOP case. The diagonal Hamiltonian for $\mathbf{k}\neq0$ close to $K_{\xi}$ is:
{
\begin{widetext}
\begin{equation}
\label{Heff22basis}
\begin{split}
H_{\rm HCP}(\mathbf{k}) &=-\mu'
\left(
\begin{array}{cccc}
_{A\uparrow+B\uparrow} & _{A\downarrow+B\downarrow} &  _{A\uparrow-B\uparrow} &  _{A\downarrow-B\downarrow} \\
1 & 0 & 0 & 0 \\
0 & 1 & 0 & 0 \\
0 & 0 & 1 & 0 \\
0 & 0 & 0 & 1
\end{array}
\right) + \frac{v\hbar}{\gamma_0}|\mathbf{k}| (h'_{0x}+\gamma_0) \left(
\begin{array}{cccc}
_{A\uparrow+B\uparrow} & _{A\downarrow+B\downarrow} &  _{A\uparrow-B\uparrow} &  _{A\downarrow-B\downarrow} \\
1 & 0 & 0 & 0 \\
0 & 1 & 0 & 0 \\
0 & 0 & -1 & 0 \\
0 & 0 & 0 & -1
\end{array}
\right)  \\
&- h'_{z0} \left(
\begin{array}{cccc}
_{A\uparrow+B\uparrow} & _{A\downarrow+B\downarrow} &  _{A\uparrow-B\uparrow} &  _{A\downarrow-B\downarrow} \\
1 & 0 & 0 & 0 \\
0 & -1 & 0 & 0 \\
0 & 0 & 1 & 0 \\
0 & 0 & 0 & -1
\end{array}
\right)  + \frac{v\hbar}{\gamma_0}|\mathbf{k}| h'_{zx} \left(
\begin{array}{cccc}
_{A\uparrow+B\uparrow} & _{A\downarrow+B\downarrow} &  _{A\uparrow-B\uparrow} &  _{A\downarrow-B\downarrow} \\
1 & 0 & 0 & 0 \\
0 & -1 & 0 & 0 \\
0 & 0 & -1 & 0 \\
0 & 0 & 0 & 1
\end{array}
\right)\\
&=-\mu' \mathbbm{1}_{\sigma} \mathbbm{1}_{s} +\frac{v\hbar}{\gamma_0}|\mathbf{k}| (h_{0x}'+\gamma_0)  \widetilde{\sigma}_{z}\mathbbm{1}_{s} -h_{z0}' \mathbbm{1}_{\sigma}S_{z}  + \frac{v\hbar}{\gamma_0}|\mathbf{k}| h_{zx}' \widetilde{\sigma}_{z}S_{z},
\end{split}
\end{equation}
\end{widetext}}                                                                                                                                                                                        
\noindent where $\widetilde{\sigma}_{z}$ is the pseudospin matrix in the basis of Eq.~(\ref{baseHeff2}). Within this basis, the interpretation of the terms is not so straight-forward as for the ATOP configuration. However, one can see that the {first} term also shifts the site energy, with a chemical potential $\mu'=0.427~$eV, that represents a transfer of electrons from Co/Ni to graphene. We can see this effect in Fig.~\ref{bandstructures} (right panels). For the {second} term we have $h_{0x}'=-0.142~$eV, and looking at Eq.~(\ref{Heff22basis}), we see that the states {$A\uparrow-B\uparrow$} and {$A\downarrow-B\downarrow$} are equally favorable, indicating symmetry between the sublattices A and B. This is because in the sublattice symmetric configuration, both A and B, are at the HCP sites of Co (Figs.~\ref{Configurations} and \ref{orbitconf2}).

{The magnetic order of graphene, with respect to Co magnetic order}, is determined by the eigenvalue of lowest energy and its corresponding eigenvector in Eq.~(\ref{Heff22basis}). Given that $h_{z0}'=0.276~$eV and $h_{zx}'=-0.092~$eV, this state corresponds to {$|A\uparrow-B\uparrow\rangle$}. 
%with eigenvalue $H_{\gamma}-\mu'+h_{0x}'-h_{z0}'+h_{zx}'$. 
Using this state, with $k_{y}=0$ and $\xi=1$, we have
\begin{eqnarray}
\label{magneticHeff2}
\langle S_{z}\rangle&=&\langle A\uparrow {-}B\uparrow|\mathbbm{1}_{\sigma}S_z|A\uparrow {-}B\uparrow\rangle\notag \\
&=& \frac{1}{2} \left(
\begin{array}{c c c c}
1 & 0 & {-}1 & 0 
\end{array}
\right)\left(
\begin{array}{c c c c}
1 & 0 & 0 & 0 \\
0 & -1 & 0 & 0 \\
0 & 0 & 1 & 0 \\
0 & 0 & 0 & -1 
\end{array}
\right)\left(
\begin{array}{c}
1 \\
0 \\
{-}1 \\
0
\end{array}
\right) = 1,\notag \\
\end{eqnarray}
so every band has a full spin polarization in either of the two spin orientations as depicted in right-hand panel of Fig.\ref{bandstructures}.

{The coupling of the spin and kinetic energy (see last term in Eq.\ref{Heff22basis}) induces a striking behavior which mimicks a spin-orbit coupling induced by the bias current and the magnetism of the Co. In the sense of equilibrium/persistent currents\cite{Bolivar}, at ${k}=0$ 
all bands have occupation below the Fermi energy, thus the spin polarization is zero at both K points. As  $k_x$ increases e.g. in the positive direction, (see Fig. \ref{bandstructures} upper-right panel) one of the bands emerges above the Fermi level and we have a net polarization which is up spin. A range of $k_x$ values preserves this polarization until a second band emerges from the Fermi sea, then the polarization returns to zero. The same behavior occurs in the opposite $k_x$ direction. This behavior is also borne out from the ATOP configuration but within a smaller wave-vector range (see Fig. \ref{bandstructures}  upper-left panel) in the vicinity of the K point. Note that this term is not derived from the atomic SOI (as is the case for both the intrinsic and Rashba interactions) but is purely parameterized by the spin-splitting energy of the Co and the wave-vector deviation from the K point. As can
be seen from Eqs.\ref{identconf1} and \ref{identconf2}, if the spin-splitting energy $\delta_{1,2}$ vanishes, this term does not appear.}

Following the lowest energy occupied states, the system is ferromagnetic in the vicinity of $k=0$. At the K point we have degenerate
bands with the same spin orientation as the Co layer, nevertheless, driving a current by means of an external electric field in the graphene
plane, one can tune the $k_x$ vector so that two oppositely oriented bands are the lowest occupied bands, making the magnetization ground state
zero.  So we have magnetic state switching controlled by the charge current on the graphene layer. 

Various scenarios of interest can be explored by using, as proposed in ref. \onlinecite{Macdonald2013}, a Cu surface so as to sandwich the graphene layer
between Co and Cu. The Cu surface will serve to control the Fermi level and access differently polarized magnetization states 
as a function of gate voltage and charge current.

\section{Conclusions}
We have derived, within the perturbative tight-binding approximation, the spectral signatures of two
Co-graphene registries, the ATOP (one Co atom atop of each A carbon atom) configuration
and the HCP (Co at the centers of the hexagonal cells of graphene). Each registry produces a very
different spectrum: a) The ATOP configuration generates a gap in pristine graphene with spin dependent
electron heavy and light effective masses for both the conduction and valence bands
that are tunable controlling orbital overlaps. As found by DFT, the graphene layer becomes almost
perfectly antiferromagnetic with down spin orientations at sublattice A and up spin favoured orientation
at sublattice B. b) The HCP configuration preserves
the linear dispersion of graphene, with a small modification of the fermion velocities. The resulting
linear dispersions shift in energy according to the spin orientation favoured on the sublattices.
For this configuration, ferromagnetic order is preferred and it is parallel to the Co polarization.
We have suggested ways to manipulate the magnetic state of the surface by  applying a gate
voltage (in the work function regime) and by driving a current through the system.
There is peculiar coupling between spin and electron momentum induced by the magnetic state of the Co. It amounts
to a spin-orbit coupling induced by the driving current. This feature is worth while
exploring in the future for both its transport and topological implications in graphene nanoribbons.

{Using Co and Ni interchangeably in this work is a good approximation as can be judged from
detailed DFT calculations\cite{DFTReference}. Nevertheless, there are some quantitative differences
in the amount of charge transfer and the magnetic moment on the graphene mainly induced by slight
changes in the bonding lengths both in the graphene and the interface layer involved. For the ATOP 
configuration the charge transfers per carbon atom are almost identical between Co and Ni, but
the induced magnetization can be two times higher for Ni for small number of layers of the metal. Also
the gap induced in the ATOP configuration can be manipulated slightly by changing the number of layers
without changing the qualitative picture. It remains to be seen what the corresponding effects are from
the HCP, configuration. }

\acknowledgements
We thank R. Kiehl for proposing we address this problem. We acknowledge the hospitality of the Chemistry
department of ASU for hosting one of us (EM) as a Fulbright Scholar. BB and EM acknowledge support 
from the project PICS-CNRS 2013-2015.

\appendix
\section{Direction cosine sums}
The summation $\sum_{m,l=1}^{3} [\hat{n}_{lx}\hat{n}_{lmx} + \hat{n}_{ly}^{2}\hat{n}_{lmy}] $ is performed as follows:
Performing first the sum over $l$ i.e. over the Co/Ni atoms $l=1,2,3$ as can be seen in Fig.~\ref{conf2}, $n_{1x}=0$, because there is no overlap 
between A and Co in $\hat{x}$ for $l=1$. The other terms are $n_{1y}=-1$, $n_{2y}=n_{3y}=1/2$ and  $n_{2x}=-\sqrt{3}/2$ $n_{3x}=\sqrt{3}/2$. 
Therefore
\begin{eqnarray}
 &&{\sum_{l=1}^{3}\sum_{m=1}^{3}(\hat{n}_{lx}\hat{n}_{lmx}+\hat{n}_{ly}\hat{n}_{lmy})}=\notag \\
 &&\sum_{m=1}^{3}[-n_{1my}+\frac{n_{2my}+n_{3my}}{2}+\frac{\sqrt{3}}{2}(n_{3mx}-n_{2mx})].\notag \\
\end{eqnarray}
Now performing the remaining sum, for $m=1$ only the cobalts $l=2$ and $l=3$ intervene, so $n_{11y}=0$, $n_{21y}=n_{31y}=1/2$ and $n_{21x}=\sqrt{3}/2$, $n_{31x}=-\sqrt{3}/2$. Doing the sum for $m=1$ we have
\begin{equation}
m=1: \Big[-0+\frac{1/2+1/2}{2}+\frac{\sqrt{3}}{2}\Big(-\frac{\sqrt{3}}{2}-\frac{\sqrt{3}}{2}\Big)\Big]=-1.\notag
\end{equation}
For $m=2$ only the Co/Ni $l=2$ and $l=1$ intervene, so $n_{12y}=1/2$, $n_{22y}=-1 $ ,$n_{32y}=0$, $n_{22x}=0$ and $n_{32x}=0$. 
Doing the sum for $m=2$ one obtains
\begin{displaymath}
m=2: \Big[-\frac{1}{2}+\frac{-1+0}{2}+\frac{\sqrt{3}}{2}\Big(0-0\Big)\Big]=-1.
\end{displaymath}
Finally, for $m=3$ the intervening Co/Ni are $l=1$ and $l=3$, so $n_{13y}=1/2$, $n_{23y}=0$, $ n_{33y}=-1$, $n_{23x}=0$ and $n_{33x}=0$. The sum for $m=3$ is then
\begin{displaymath}
m=3: \Big[-\frac{1}{2}+\frac{0-1}{2}+\frac{\sqrt{3}}{2}\Big(0-0\Big)\Big]=-1.
\end{displaymath}
So in spite of the complicated combination of direction cosines, all the matrix overlaps are equivalent.
\section{Parameter values of the ATOP Hamiltonian}
From Eq.~\ref{identconf1}, with $\varepsilon=\varepsilon_p-\varepsilon_d$, one obtains
%\begin{widetext}
\begin{eqnarray}
\mu&=&-\frac{\varepsilon}{2}\left(\frac{V^2_{pdz}}{\varepsilon^{2}-\delta_{1}^{2}}+3\frac{\widetilde{V}^2_{pd\pi}}{\varepsilon^{2}-\delta_{2}^{2}}\right), \\
h_{0z}&=&\frac{\varepsilon}{2}\left(\frac{V^2_{pdz}}{\varepsilon^{2}-\delta_{1}^{2}}-3\frac{\widetilde{V}^2_{pd\pi}}{\varepsilon^{2}-\delta_{2}^{2}}\right), \\
h_{z0}&=&\frac{\delta_1V^2_{pdz}}{\varepsilon^2-\delta_{1}^{2}} +3\frac{\delta_2\widetilde{V}^2_{pd\pi}}{\varepsilon^2-\delta_{2}^{2}},\\
h_{zz}&=&\frac{\delta_1V^2_{pdz}}{\varepsilon^2-\delta_{1}^{2}} -3\frac{\delta_2\widetilde{V}^2_{pd\pi}}{\varepsilon^2-\delta_{2}^{2}}.
\end{eqnarray}
%\end{widetext}

\end{document}